\documentclass[journal]{IEEEtran}

\ifCLASSINFOpdf
\else
   \usepackage[dvips]{graphicx}
\fi
\usepackage{url}
\usepackage{graphicx}
\usepackage{color}
\usepackage{bm}
\usepackage{float}
\usepackage{soul}
\usepackage{xcolor}
\usepackage{amsmath}
\usepackage{color}
\usepackage{bm}
\usepackage{float}
\usepackage{tabularx}
\usepackage{gensymb}
\usepackage[numbers,sort&compress]{natbib}

\begin{document}

\title{Spatial-Filter-Bank-Based Neural Method for Multichannel Speech Enhancement}

\author{\it Tianqin Zheng, Jilu Jin, Hanchen Pei, Gongping Huang, Jingdong Chen, and Jacob Benesty}

\markboth{IEEE SIGNAL PROCESSING LETTERS, VOL. XX, NO. XX, JAN 2025}
{Shell \MakeLowercase{\textit{et al.}}: Bare Demo of IEEEtran.cls for IEEE Journals}
\maketitle

\begin{abstract}
The performance of deep learning-based multichannel speech enhancement methods often deteriorates when the geometric parameters of the microphone array change. Traditional approaches to mitigate this issue typically involve training on multiple microphone arrays, which can be costly. To address this challenge, we focus on uniform circular arrays and propose the use of a spatial filter bank to extract features that are approximately invariant to geometric parameters. These features are then processed by a two-stage conformer-based model (TSCBM) to enhance speech quality. Experimental results demonstrate that our proposed method can be trained on a fixed microphone array while maintaining effective performance across uniform circular arrays with unseen geometric configurations during applications. 
\end{abstract}


\IEEEpeerreviewmaketitle

\section{Introduction}
\label{sec:intro}

Speech enhancement techniques aim to improve the quality and/or intelligibility of speech signals. Many deep learning-based architectures have been developed for this purpose~\cite{Tan2018ACR, cao2022cmgan, schroter2022deepfilternet, kim2021se, yu2022dual, fu2019metricgan}, delivering promising results. However, most of these methods are designed for single-channel scenarios, overlooking the important spatial information in more complex environments. As a result, multichannel enhancement approaches have been developed~\cite{liu2020multichannel, pop00001, zhang2021adl, casebeer2021nice, 8461639, tolooshams2020channel}, though these are often tailored to specific microphone array topologies, leading to suboptimal performance when applied to unseen array geometries. This limitation stems from the models' inability to adapt to changes in array geometry. Therefore, achieving model generalization across diverse array setups is crucial for reducing the need for dataset reconstruction and retraining.

Some methods have been proposed to improve the adaptability of the model to address the challenge mentioned above \cite{luo2020end, pandey2022tparn, zhang2023toward, lin2024agadir, yoshioka2022vararray}, including applying targeted-acceleration-and-compression (TAC) layers to optimize microphone data integration \cite{luo2020end}, extracting inter-channel-phase-difference (IPD) features to acquire spatial information \cite{yoshioka2022vararray}, and building a triple-path network based on spatial self-attention to process array observations \cite{pandey2022tparn}. Although these methods are effective, they face notable restrictions: they all require separate processing for each channel, which leads to high computational overhead, and they rely on diverse datasets for generalization, which is not feasible for real-world implementations.

To overcome these challenges, we focus on training a model using a fixed array to ensure robust performance across various geometries. We introduce a spatial filter bank for feature extraction that remains nearly invariant to geometric variations. For simplicity, we use uniform circular arrays (UCAs) to develop the proposed algorithms. By utilizing a model based on~\cite{cao2022cmgan}, we process these features to enhance the speech signal, simultaneously reducing computational complexity through joint channel processing. Experimental results demonstrate that our approach outperforms existing methods and maintains strong performance on unseen arrays.

The key contributions of this work are threefold. First, we identify the critical factors influencing model performance across different microphone arrays and propose an interpretable feature extraction method to ensure consistent high performance. Second, the proposed method is highly versatile, capable of integrating with various multichannel speech enhancement models to improve their generalization to unseen arrays. Third, while our study primarily focuses on UCAs for simplicity, the feature extraction technique can be extended to arrays with arbitrary geometries, e.g., the ones discussed in \cite{huangarbitrary}.

\section{Signal Model and Problem Formulation}
\label{sec:signal model}

Consider a UCA consisting of $M$ omnidirectional sensors uniformly spaced around a circle of radius $r$.  Taking the center of the UCA as the reference point and assuming a plane wave arriving from an azimuth angle $\theta$, the phase delay at the $m$th sensor relative to the reference can be expressed as $\zeta_m \left( \omega, \theta \right) = e^{\jmath \overline{\omega} \cos \left( \theta - \psi_m \right)}$, $m = 1, 2, \ldots, M$, where $\jmath$ is the imaginary unit, $\overline{\omega} = \omega r / c$, $\omega$ is the angular frequency, and $c$ denotes the speed of sound in air. The steering vector can then be represented as
\begin{align}
    \label{sv}
    \mathbf{d} \left( \omega, \theta \right)
    & = \left[ \begin{array}{cccc}
    \zeta_1 \left( \omega, \theta \right) & \zeta_2 \left( \omega, \theta \right) & \cdots & \zeta_M \left( \omega, \theta \right)
    \end{array}\right]^T,
\end{align}
where the superscript $^T$ is the transpose operator. In the short-time-Fourier-transform (STFT) domain, the signals received by the UCA can be written as
\begin{align}
    \label{y-vect-def}
    \mathbf{y} \left( k,\omega \right)
    & = \left[ \begin{array}{cccc}
    Y_1 \left( k, \omega \right) & Y_2 \left( k, \omega \right) & \cdots & Y_M \left( k, \omega \right)
    \end{array} \right]^T \nonumber \\
    & = \mathbf{x}_\mathrm{d} \left( k, \omega \right) + \mathbf{v} \left( k, \omega \right),
\end{align}
where $\mathbf{v} \left( k,\omega \right)$, of length $M$, is the vector of the background noise, and
\begin{align}
    \label{x-d-vect-def}
    \mathbf{x}_\mathrm{d} \left( k, \omega \right)
    & = \left[ \begin{array}{ccc}
    X_{1, \mathrm{d}} \left( k, \omega \right) & \cdots & X_{M, \mathrm{d}} \left( k, \omega \right)
    \end{array} \right]^T,
\end{align}
also of length $M$, represents the desired signal originated from the source of interest,  with (see Subsection~\ref{ssec:SFE} for more details)
\begin{align}
    \label{x-d-def}
    X_\mathrm{d} \left( k, \omega \right)
    & = G \left( \omega \right) S \left( k, \omega \right),
\end{align}
$G \left( \omega \right)$ being the transfer function corresponding to the direct path and early reflections,  and $S \left( k, \omega \right)$ denoting the source signal. 

The objective of this work is to estimate $X_\mathrm{d} \left( k, \omega \right)$ from the array observed vector, $\mathbf{y} \left( k, \omega \right)$. To accomplish this, we derive a neural network (NN) based method.

\begin{figure}[t!]
    \centering
    \includegraphics[width=0.475\textwidth]{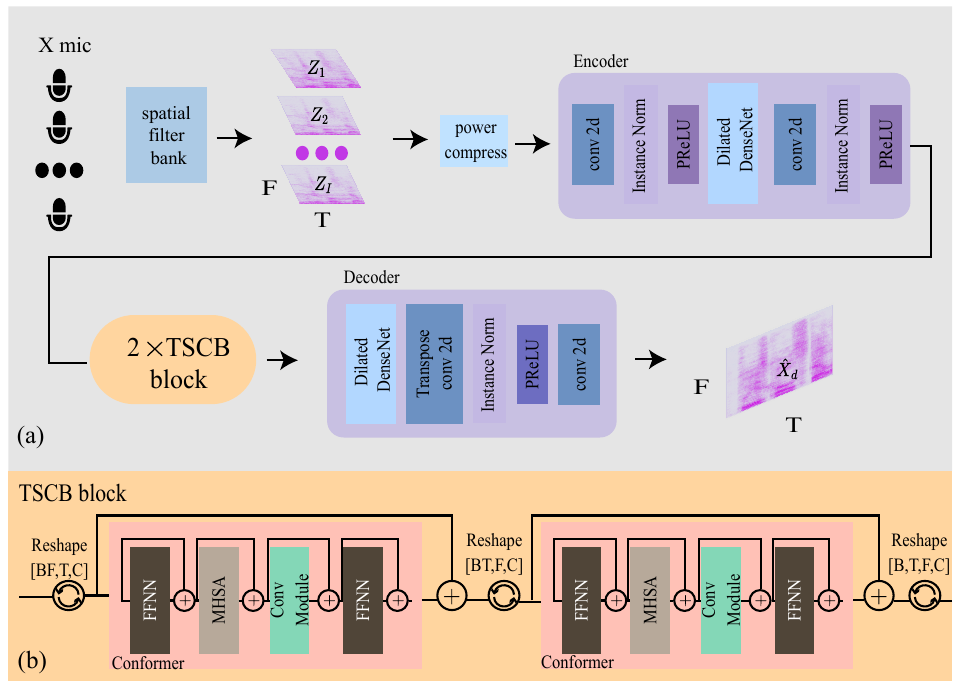} 
    \caption{The overall architecture of the proposed TSCBM+FB.}
    \label{fig:whole}
\end{figure}

\section{Method Derivation}
\label{sec:method}

To address the performance degradation of trained models on previously unseen arrays, we introduce a method called TSCBM+FB, which stands for two-stage conformer-based model (TSCBM) with filter bank (FB). This approach first extracts features that are independent of the array radius and the number of microphones using a spatial FB. These features are then processed and enhanced by the TSCBM. The overall architecture of this model is illustrated in Fig.~\ref{fig:whole}. In the following subsections, we detail the feature extraction process and provide a comprehensive explanation of the TSCBM architecture.

\subsection{Spatial Feature Extraction}
\label{ssec:SFE}

Given the decomposition of the spatial room impulse responses, the transfer function, $G \left( \omega \right)$, is composed of contributions from a set of image sources; consequently, $G \left( \omega \right)$ can be expressed as follows:
\begin{align}
    \label{g-fact}
    G \left( \omega \right)
    & = \sum^L_{l = 1} Q_l \left( \omega \right) ,
\end{align}
where $Q_l \left( \omega \right)$ represents the spectrum of the impulse response corresponding to the $l$th image source with $L$ being the total number of image sources. The desired component of the source signal received by the $m$th microphone can then be written as
\begin{align}
    \label{x-m-d-fact}
    X_{m, \mathrm{d}} \left( k, \omega \right)
    & = \sum^L_{l = 1} Q_l \left( \omega \right) \zeta_m \left( \omega, \theta_l \right) S \left( k, \omega \right),
\end{align}
where $\zeta_m \left( \omega, \theta_l \right)$ denotes the phase delay of the $l$th path to the $m$th microphone. The vector $\mathbf{x}_\mathrm{d} \left( k, \omega \right)$ can then be expressed as
\begin{align}
    \label{x-d-vect-def-2}
    \mathbf{x}_\mathrm{d} \left( k, \omega \right)
    & = \sum^L_{l = 1} Q_l \left( \omega \right) S \left( k, \omega \right) \mathbf{d} \left( \omega, \theta_l \right).
\end{align}
As indicated by \eqref{x-d-vect-def-2}, changes in the UCA geometry, such as changes in the number of sensors or the radius, will affect $\mathbf{d} \left( \omega, \theta_l \right)$. This, in turn, can lead to performance degradation in deep learning models. Therefore, it is crucial to extract features that are independent of the array geometry. Specifically, a spatial filter, $\mathbf{h} \left( \omega \right)$, can be used to achieve this. Then, the filtered signal is
\begin{align}
    \label{z-def}
    Z \left( k, \omega \right)
    & = \mathbf{h}^H \left( \omega \right) \mathbf{y} \left( k, \omega \right) \nonumber \\
    & = \mathbf{h}^H \left( \omega \right) \mathbf{x}_\mathrm{d} \left( k, \omega \right) + \mathbf{h}^H \left( \omega \right) \mathbf{v} \left( k, \omega \right) \nonumber \\
    & = X_\mathrm{fd} \left( k, \omega \right) + V_\mathrm{rn} \left( k, \omega \right),
\end{align}
where the superscript $^H$ is the conjugate-transpose operator, $X_\mathrm{fd} \left( k, \omega \right)$ is the filtered desired signal, and $V_\mathrm{rn} \left( k, \omega \right)$ is the residual noise. 

Let us examine the structure of $X_\mathrm{fd} \left( k, \omega \right)$, which is
\begin{align}
    \label{x-fd-def}
    X_\mathrm{fd} \left( k, \omega \right)
    & = \sum^L_{l = 1} Q_l \left( \omega \right) S \left( k, \omega \right) \mathbf{h}^H \left( \omega \right) \mathbf{d} \left( \omega, \theta_l \right) \nonumber \\
    & = \sum^L_{l = 1} Q_l \left( \omega \right) S \left( k, \omega \right) \mathcal{B} \left[ \mathbf{h} \left( \omega \right), \theta_l \right],
\end{align}
where $\mathcal{B} \left[ \mathbf{h} \left( \omega \right), \theta_l \right]$ denotes the spatial response, i.e., the beampattern at azimuth angle $\theta_l$. 
As shown in \eqref{x-fd-def}, $Q_l \left( \omega \right)$ and $S \left( k, \omega \right)$ are not dependent on the array geometry. If the spatial response, $\mathcal{B} \left[ \mathbf{h} \left( \omega \right), \theta_l \right]$, of the spatial filter is independent of the geometry of the UCA, then the extracted speech feature will also be independent of the geometric parameters.

To ensure that the beampattern remains independent of the geometric parameters, we can design the filter to approximate the desired directivity pattern. To achieve this, we employ the method proposed in~\cite{huang2017design}. For a UCA, the ideal beampattern with the mainlobe directed towards $\theta_\mathrm{s}$ can be expressed as
\begin{align}
    \mathcal{B} \left( \mathbf{b}_{N}, \theta \right)
    & = \sum_{n=-N}^N b_{N,n} e^{\jmath n \left( \theta - \theta_\mathrm{s} \right)} \nonumber \\
    & = \mathbf{b}_N^T \mathbf{p} \left( \theta \right),
\end{align}
where $\mathbf{b}_N$ contains the coefficients determining the shape of the ideal beampattern, $N$ is the order of the beampattern, and 
\begin{align}
    \mathbf{p} \left( \theta \right)
    & = \left[\begin{array}{ccccc}
    e^{-\jmath N \theta} & \cdots & 1 & \cdots & e^{\jmath N \theta}
    \end{array} \right]^T.
\end{align}
The design of the beamforming filter can be approached from a least-squares perspective \cite{huang2017design}. The filter is formulated as follows:
\begin{align}
    \label{h-vect-def}
    \mathbf{h} \left( \omega \right)
    & = \frac{1}{M} \mathbf{\Psi}^H \mathbf{J}^* \left( \overline{\omega} \right) \mathbf{\Upsilon}^* \left( \theta_{\mathrm{s}} \right) \mathbf{b}_N,
\end{align}
where the superscript $^*$ is the conjugate operator, and
\begin{align}
    \mathbf{\Psi}
    & = \left[ \begin{array}{ccccc}
    \boldsymbol{\psi}_{-N}^* & \cdots & \boldsymbol{\psi}_0^* & \cdots & \bm{\psi}_N^*
    \end{array} \right]^T, \\
    \boldsymbol{\psi}_n
    & = \left[ \begin{array}{cccc}
    e^{-\jmath n \psi_1} & e^{-\jmath n \psi_2} & \cdots & e^{-\jmath n \psi_M}
    \end{array} \right]^T, \\
    \mathbf{J} \left( \overline{\omega} \right)
    & = \mathrm{diag} \left[ \frac{1}{\jmath^{-N}J_{-N} \left( \overline{\omega} \right)} \cdots \frac{1}{J_0 \left( \overline{\omega} \right)} \cdots \frac{1}{\jmath^N J_N \left( \overline{\omega} \right)} \right], \\
    \mathbf{\Upsilon} \left( \theta_\mathrm{s} \right)
    & = \mathrm{diag} \left[ e^{\jmath N \theta_\mathrm{s}} ~~\cdots~~ 1 ~~\cdots~~ e^{-\jmath N \theta_\mathrm{s}} \right],
\end{align}
with $J_n \left( \cdot \right)$ being the $n$th-order Bessel function of the first kind. 

\begin{figure}[htbp]
\centering
\includegraphics[width=0.3\textwidth]{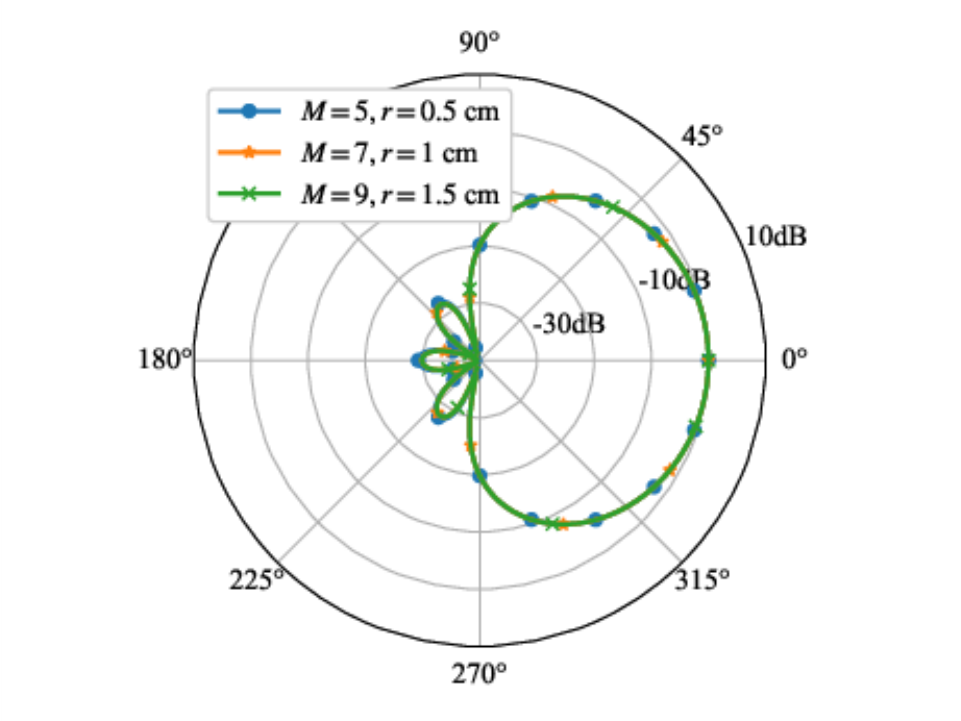} \vskip -12pt
\caption{Designed second-order supercardioid beampattern with varying UCA geometric parameters. Conditions: $\theta_\mathrm{s} = 0^\circ$ and $f = 4$~kHz.}
\label{fig:beampattern1}
\end{figure}

Figure~\ref{fig:beampattern1} shows the beampatterns of a supercardioid beamformer designed for a UCA with different geometric parameters. The beampatterns are highly consistent, demonstrating that they remain largely unaffected by geometric variations. This consistency highlights that the features extracted using this method are primarily independent of geometric parameters, which is crucial for maintaining robustness across various array configurations.

Although the beampattern of a spatial filter can be designed to be independent of the geometric parameters of the UCA, it mainly enhances sound from a specific direction, $\theta_\mathrm{s}$, while attenuating sound from other directions, as shown in Fig.~\ref{fig:beampattern1}. Since the speaker’s location is typically unknown in practice, it is necessary to use a set of spatial filters oriented in different directions to form a spatial FB for feature extraction. Specifically, we use a total of $I$ filters, where the $i$th filter orients toward $\theta_\mathrm{s} = \frac{i}{I} 2\pi$, and the output from this filter is denoted as $Z_i$.

\subsection{Model Architecture}

Our network is built upon CMGAN \cite{cao2022cmgan}, which incorporates a well-established dual-path architecture that effectively captures both temporal and frequency information in speech signals, demonstrating outstanding performance in speech enhancement tasks. To address the specific challenges we encountered, we streamlined the network structure and adapted the proposed approach to the TSCBM. It is important to highlight that our method is not limited to TSCBM and can be applied to most multichannel speech enhancement models. We choose TSCBM as the backbone model to simultaneously process multichannel signals, eliminating the need for separate processing of individual channels \cite{luo2020end, pandey2022tparn, yoshioka2022vararray}, which is crucial for reducing computational complexity.

\subsubsection{Input Features}

The spatial FB's output features are compressed to equalize the significance of softer sounds against louder ones:
\begin{align} 
Z_i^{'} = |Z_i|^c e^{\jmath \angle{Z_i}},
\end{align}
where $Z_i^{'}$ is the compressed feature and $c=0.3$ is the compression exponent as per \cite{cao2022cmgan}. The real and imaginary parts of $Z_i^{'}$ are concatenated to form the input $Z^{'} \in \mathcal{R}^{B \times 2I \times T \times F}$, with $T$ and $F$ representing time and frequency dimensions, respectively.

\subsubsection{Encoder}
The encoder maps $Z^{'}$ into a latent feature space. It initiates with a convolutional block with a kernel of $(1,1)$ and stride $(1,1)$, followed by instance normalization and PReLU activation. This yields an intermediate feature map $[B,C,T,F]$ with $C=64$. A dilated DenseNet with dilation factors of 1 and 2 is then applied. The process ends with a convolutional block with a kernel of $(1,3)$ and stride $(1,2)$, downsampling the frequency dimension.

\subsubsection{TSCB Block}
Intermediate features pass through two-stage conformer blocks (TSCBs) to capture temporal and frequency dependencies. Each TSCB, as shown in Fig.~\ref{fig:whole}(b), contains two conformer blocks, each addressing temporal and frequency dependencies. These blocks include two FFNNs, an MHSA mechanism with four heads, and a convolution module. Skip connections are used to preserve feature integrity.

\subsubsection{Decoder}
The decoder reconstructs the desired signal spectrum. It begins with a dilated DenseNet mirroring the encoder's architecture. A sub-pixel convolution block follows, doubling the feature dimension to $C=128$ and upsampling the frequency dimension via pixel shuffle. The final convolution block includes instance normalization, PReLU activation, and a kernel of $(1,2)$, yielding a final feature map with $C=2$ for the real and imaginary parts of the signal spectrum.

\section{Experiments}
\label{sec:exp}
\vspace{10pt}

\subsection{Experimental Setup}
\vspace{10pt}
\subsubsection{Dataset}

We simulated a multichannel dataset using the VoiceBank and DEMAND datasets. The training set features clean speech mixed with 12 types of background noise from DEMAND at SNR levels ranging from $-5$ to $10$~dB. The test set introduces 5 new noise types from DEMAND, with multichannel RIR generated via the image model. Room dimensions vary from $3$~m $\times$ $3$~m $\times$ $2.5$~m to $7$~m $\times$ $9$~m $\times$ $3$~m, and reverberation time ($T_{60}$) is randomly set between $0.2$ and $0.35$ seconds. Microphone array, sound source, and noise source positions are randomly determined, with a minimum directional angle difference of $5^\circ$ between the target speech and interference. To test generalization across array geometries, the number of microphones and UCA radius differ between training (5 microphones, 0.5~cm radius) and test sets (7 or 9 microphones, 1~cm or 1.5~cm radius). Unlike most related work that trains on multiple arrays for generalizability, our approach trains on a single UCA type and evaluates performance across four distinct UCA types.

Evaluation metrics include floating point operations per second (FLOPS) of the model, perceptual evaluation of speech quality (PESQ)~\cite{pesq},
mean opinion score (MOS) predictions of speech distortion (CSIG), MOS predictions of intrusiveness of background noise (CBAK), MOS predictions of overall processed speech quality (COVL)~\cite{csigxx}, and STOI~\cite{stoi}. PESQ ranges from $-0.5$ to $4.5$, CSIG, CBAK, and COVL from $1$ to $5$, and STOI from $0$ to $1$. Apart from FLOPS, higher scores on all other metrics indicate better performance.

\subsubsection{Training Configuration}
The training set utterances are truncated into $2$-second segments, while the test set uses full-length sentences. Following the method outlined in~\cite{cao2022cmgan}, we apply a Hamming window with a $25$-ms window length (corresponding to 400-point FFT) and a hop size of $6.25$~ms (i.e., $75\%$ overlap).
A total of 9, (i.e., $I=9$) spatial filters are used, which are oriented at $\theta_\mathrm{s} = \frac{i}{I} 2 \pi$ to extract $I$ features, employing a second-order supercardioid filter. The coefficients for the ideal beampattern used to design this filter are $\mathbf{b}_N = \left[0.1035~ 0.242~ 0.309~ 0.242~ 0.1035\right]^T$.
The loss function is based on the mean-squared error of the real part, imaginary part, and magnitude of the estimated spectrum. During training, the AdamW optimizer~\cite{loshchilov2017decoupled} is employed with a learning rate of $5 \times 10^{-4}$.

\subsubsection{Comparison Methods}
For comparison, we employ the FaSNet+TAC network~\cite{luo2020end} as a baseline. Originally designed for permutation- and number-invariant speech separation, it is modified for speech enhancement by omitting the beamforming component due to the absence of a reference center microphone. This modified model is termed DPRNN+TAC. 
Furthermore, we introduce a DPRNN+FB model, a variant of DPRNN+TAC that removes the TAC module, enabling joint channel processing. This model is integrated with our feature extraction method to show that our approach can effectively avoid the high computational complexity typically associated with TAC-like modules in array geometry-agnostic tasks.
To address the degradation of deep learning models on unseen UCAs, we train TSCBM on a fixed array with 5 microphones and a radius $0.5$~cm. For testing, if the test array exceeds 5 microphones, we select $5$ with azimuth angles closest to the training array as the TSCBM input, referred to as TSCBM+select. Our proposed method, incorporating spatial FB, is dubbed TSCBM+FB.

\subsection{Experimental Results}
\begin{table}[htbp]
\centering
 \caption{Speech enhancement performance of the DPRNN+TAC, TSCBM+select, and  TSCBM+FB methods.}
 \renewcommand{\arraystretch}{1.2}
\scalebox{1}{
\setlength{\tabcolsep}{5pt}
\begin{tabular}{lcccccc} 
\hline
Model     &FLOPS   & PESQ          & CSIG          & CBAK          & COVL          & STOI           \\ \hline
Noisy      &\ & 1.23          & 2.72          & 1.86          & 1.99          & 0.789           \\ \hline
DPRNN+TAC  &31.3G    & 2.25          & 4.09          & 3.06          & 3.23          & 0.914          \\
TSCBM+select &28.3G    & 2.50          & 4.02          & 3.21          & 3.32          & 0.934          \\
DPRNN+FB &\textbf{2.77G}     &2.25 &4.15 &2.98 &3.26 &0.911\\
TSCBM+FB    &28.3G    &\textbf{2.76} & \textbf{4.36} & \textbf{3.30} & \textbf{3.64} & \textbf{0.947} \\ 
\hline
\end{tabular}}
\label{tab:table1}
\end{table}
We first evaluate the performance of these methods on the test set. The results are presented in Table~\ref{tab:table1} in details; they indicate that TSCBM+FB excels when our feature extraction methods are employed, underscoring the model's ability to adapt to unseen microphone arrays by extracting geometry-independent features. It is seen that DPRNN+FB and DPRNN+TAC exhibit comparable performance, even though the computational complexity of DPRNN+FB is approximately one-tenth that of DPRNN+TAC. This indicates that our feature extraction methods can substantially reduce computational demands in array geometry-agnostic tasks. Furthermore, TSCBM+select outperforms DPRNN+TAC, validating the effectiveness of selecting microphones similar to those used in training. 
\begin{figure}[t!]
	\centering
	\begin{minipage}{0.42\linewidth}
		\centering
		\includegraphics[width=0.9\linewidth]{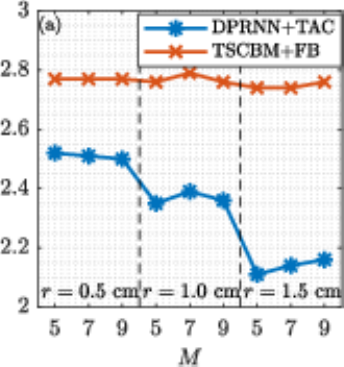}
	\end{minipage}
	\begin{minipage}{0.42\linewidth}
		\centering
		\includegraphics[width=0.9\linewidth]{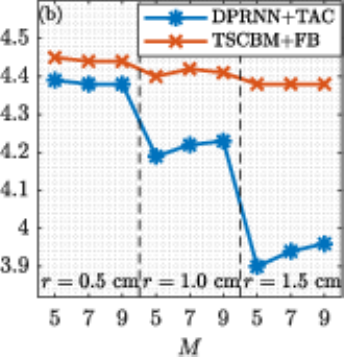}
	\end{minipage}
	
	\begin{minipage}{0.42\linewidth}
		\centering
		\includegraphics[width=0.9\linewidth]{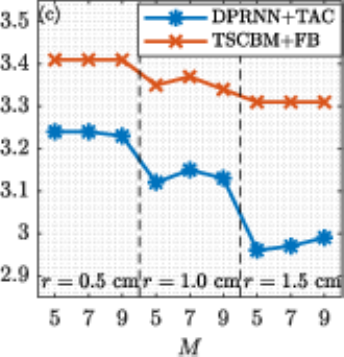}
	\end{minipage}
	\begin{minipage}{0.42\linewidth}
		\centering
		\includegraphics[width=0.9\linewidth]{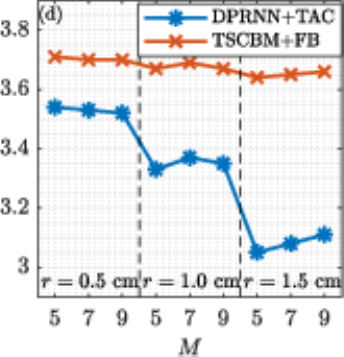}
	\end{minipage} \vskip -6pt 
 \caption{Performance comparison between DPRNN+TAC and TSCBM+FB versus different numbers of microphones and radii: (a)~PESQ, (b)~CSIG, (c)~CBAK, and (d)~COVL.}
 \label{fig:fig3}
\end{figure}
 
To evaluate the generalization of our method across microphone array geometries, we create 9 test sets with varying radii (0.5, 1, and 1.5 cm) and microphone counts (5, 7, and 9). Both DPRNN+TAC and TSCBM+FB are trained on a UCA with 5 microphones and a 0.5 cm radius. The test set configuration ($M=5$, $r=0.5$~cm) matches the training set, serving as a ``reference performance.'' Figure~\ref{fig:fig3} displays the performance metrics. DPRNN+TAC maintains consistent performance across different microphone numbers but degrades with changes in array radius, as the TAC module handles channel count variations but not radius changes. In contrast, TSCBM+FB shows stable performance across both radius and microphone count variations, indicating its robust generalization across array geometries.

\section{Conclusions}
\label{sec:conclusion}

This paper addresses the challenge of training a model on a fixed uniform circular microphone array to ensure consistent performance across various UCAs. We employ a spatial filter bank to extract geometry-independent features for speech enhancement using TSCBM. Despite its simplicity, our feature extraction method is highly effective. Our approach demonstrates robust performance and generalization across different array geometries. While we focus on circular arrays for simplicity, the underlying principles are applicable to arbitrarily shaped planar arrays, positioning our method as a potential standard for generalizing across diverse microphone configurations. 


\end{document}